# Hydrogen super-saturated layers in H/D plasma loaded tungsten: a global model based on Thermodynamics, Kinetics and Density Functional Theory data

E. A. Hodille, N. Fernandez, Z. A. Piazza, M. Ajmalghan, and Y. Ferro[*]

*Laboratoire PIIM, Aix-Marseille Université/CNRS, Avenue escadrille Normandie-Niemen, 13397, Marseille, France*

**Abstract**

In this work, we combine Density Functional Theory data with a Thermodynamic and a kinetic model to determine the total concentration of hydrogen implanted in the sub-surface of tungsten exposed to a hydrogen flux. The sub-surface hydrogen concentration is calculated given a flux of hydrogen, a temperature of implantation, and the energy of the incoming hydrogen ions as independent variables. This global model is built step by step; an equilibrium between atomic hydrogen within bulk tungsten and a molecular hydrogen gas phase is first considered, and the calculated solubility is compared with experimental results. Subsequently, a kinetic model is used to determine the chemical potential for hydrogen in the sub-surface of tungsten. Combining both these models, two regimes are established in which hydrogen is preferentially trapped at either interstitial sites or in vacancies. We deduce from our model that the existence of these two regimes is driven by the temperature of the implanted tungsten sample; above a threshold or transition temperature is the *interstitial* regime, below is the *vacancy* regime in which super-saturated layers form within tenths of angstrom below the surface. A simple analytical expression is derived for the co-existence of the two regimes depending on the implantation temperature, the incident energy and the flux of the hydrogen ions which we use to plot the corresponding phase diagram.

*corresponding author: yves.ferro@univ-amu.fr

1. **Introduction**

The solubility of hydrogen in tungsten was experimentally established by Frauenfelder [1]; in the case of a tungsten sample in thermodynamic equilibrium with a hydrogen atmosphere, the solubility is around $10^{-18}$ *at.fr.* at room temperature and under standard pressure. Under low-energy hydrogen plasma exposure, the total amount of trapped hydrogen can rise to $10^{-2}$ *at.fr.* [2] or above [2-4]. Such a high solubility is reached despite the fact that the kinetic energy of the implanted ion is below the displacement threshold of the tungsten atoms, meaning that no radiation-defects are created to accommodate additional hydrogen atoms in tungsten. Consequently, the question of how the solubility can be increased by sixteen orders of magnitude under plasma exposure constitutes the central focus of the present paper.

It is known since the work of Fukai [5] that vacancies in metal can be created under a high pressure of hydrogen in the range of several GPa. On Pd and Ni, up to 20 *at.*% of vacancies were observed and this phenomena was consequently named Super-Abundant Vacancies (SAV). Each vacancy can individually accommodate one or multiple hydrogen atoms; they constitute traps for hydrogen and induce a dramatic increase of the solubility of hydrogen. Some thermodynamic models have been developed [6-11] since then to understand the formation of SAV in metals. The driving mechanism is that hydrogen decreases the formation energy of vacancies in the host metal, which results in the formation of a huge number of vacancies above a given pressure or chemical potential of hydrogen.

Such thermodynamic models have also been developed for tungsten [12-15]. The one of Sun et al. [12] is probably the most comprehensive; it gives the number of vacancies and the total solubility of hydrogen at a given chemical potential imposed by a $H_2$ atmosphere. Based on the previous work of Sugimoto [16], Sun et. al. [12] were able to relate the chemical potential to the pressure of the gaseous hydrogen far beyond ideal gas ranges up to GPas. However, the work of [12] does not take into account the temperature dependency of the many different trapping energies and entropies of hydrogen in tungsten. In a previous work from us [13], the temperature dependency was introduced by including the vibrational energies and entropies for H in its multiple environments in tungsten: as interstitial ($H_i$), or in single vacancy (V) in which *j* H atoms can be trapped to form a $VH_j$ vacancy with *j*=0-12.

During plasma exposure however, tungsten is not in contact with molecular hydrogen, but is exposed to a flux of hydrogen ions. As a consequence, the relation between the flux/energy of the plasma particles and the chemical potential of hydrogen must be established in order to predict the amount of hydrogen retained and the number of vacancies created into the tungsten material under various exposure conditions. To this end, we use the DFT data we recently published in [13] along with an improved thermodynamic model that includes the chemical potential of hydrogen. As a consequence, the equilibrium of the tungsten sample with a hydrogen reservoir of chemical potential µ can be described. In addition to this, a kinetic model recently proposed by Schmid *et al.* [3,17] is used to take into account the dependency of the flux and the ion energy of the particles during plasma exposure. Assuming a steady state is reached, a flux balance is established between diffusion into the sample of the implanted H/D atoms and outgassing from the sample. Combining both the results of the thermodynamic and kinetic model, we are able to determine a chemical potential from the ion energy and the flux of particles.

In the end, this global model allows to determine, within the implantation depth of tungsten, the atomic fractions of interstitial hydrogen, the overall concentration in vacancies and hydrogen trapped in vacancies under hydrogen exposure at a given temperature, flux and ion energy. The existence of two regimes is established: in a first one, hydrogen is trapped at interstitial sites, while in the second regime SAV are formed, which leads to hydrogen super-saturated layers (SSL). These SSL contain atomic fraction of hydrogen and vacancies in the range of the 1% at. fr., which is consistent with the experimental observations [2-4].

## 2. DFT data

The data needed to build the thermodynamic model were previously computed in ref [13] by means of electronic structure calculations using DFT with the Quantum Espresso code [18]. Based on these data, we calculated the formation energy of an interstitial hydrogen atom $e_{int}$ in its lowest energy configuration (this occurs when hydrogen is located at tetrahedral position of the *bcc* unit cell of perfect tungsten), the formation energy of a single empty vacancy $e_0$, and the formation energy of a single vacancy $VH_j$ that traps *j* hydrogen atoms $e_j$:

$$e_{int} = E_{HW_n}^{DFT} - E_{W_n}^{DFT} - E_{H_{ref}}^{DFT} \tag{1}$$

$$e_0 = E_V^{DFT} - \frac{n-1}{n} E_{W_n}^{DFT} \tag{2}$$

$$e_j = e_V + E_{VH_j}^{DFT} - E_V^{DFT} - j\, E_{H_{ref}}^{DFT} \tag{3}$$

where $E_{W_n}^{DFT}$ is the electronic energy of a unit-cell containing n=54 atoms in ref [13], $E_{HW_n}^{DFT}$ is the energy of the same unit-cell with a hydrogen atom in a tetrahedral position, $E_V^{DFT}$ is the energy of the single vacancy (i.e. the unit-cell with one W atom removed), $E_{VH_j}^{DFT}$ is the energy of a single vacancy incorporating *j* H atoms, with *j=1* to *12*, and $E_{H_{ref}}$ is a reference energy of hydrogen. $E_{H_{ref}}^{DFT}$ can be chosen as the energy of atomic hydrogen or half the energy of a hydrogen molecule. As we are only dealing with energy difference, this reference energy will cancel out anyway. We nevertheless used half the energy of a $H_2$ molecule to calculate $e_{int}$, $e_0$ and $e_j$ displayed in Table 1. These data are in good agreement with some other ones that can be found in the literature [19-21]. For $j = 1$ to 6, the formation energy of a $VH_j$ vacancy requires less energy than the formation of an empty vacancy $e_0$; this energy difference is indeed the driving mechanism that leads to the formation of abundant $VH_j$ vacancies.

| Sites | Energies (eV) |
|---|---|
| $e_{int}$ | 0.93 |
| $e_0$ | 3.25 |
| $e_1$ | 2.99 |
| $e_2$ | 2.73 |
| $e_3$ | 2.57 |
| $e_4$ | 2.54 |
| $e_5$ | 2.57 |
| $e_6$ | 2.83 |
| $e_7$ | 3.48 |
| $e_8$ | 4.08 |
| $e_9$ | 4.81 |
| $e_{10}$ | 5.60 |
| $e_{11}$ | 6.58 |
| $e_{12}$ | 7.20 |

**Table 1:** *formation energies $e_{int}$ of an interstitial atom, $e_0$ of an empty single vacancy, and $e_j$ of a single-vacancy filled with j H atoms as calculated from equations 1, 2 and 3.*

The Gibbs free energy was further computed using the phonon properties of hydrogen in tungsten calculated via Density Functional Perturbation Theory (DFPT) [22] while keeping all the tungsten atoms frozen. This was also previously done in Ref [13] and was proven to be valid due to the mass difference of hydrogen and tungsten. The Gibbs free energy *per* particles $g_{int}$, and $g_j$ were computed with following equation:

$$g = \left(e^{DFT} + h^{vib}\right) - Ts^{vib} \qquad (4)$$

The formula for $h^{vib}$ and $s^{vib}$ are given in the Appendix. $g_V$ is simply $e_V$ since we neglected the phonon properties of the tungsten network. The standard free energy *per* particle of $H_2$ in the gas phase was also calculated, which requires adding the translational and rotational components of enthalpy and entropy to the equation (4). At standard pressure $P^\circ$, $g_{H_2}^\circ$ is determined according to:

$$g°_{H_2} = (e^{DFT}_{H_2} + h^{vib}_{H_2} + h^{rot}_{H_2} + h^{trans}_{H_2}) - T(s^{vib}_{H_2} + s^{rot}_{H_2} + s°^{trans}_{H_2}) \quad (5)$$

Again, the translation and rotational component to the chemical potential are given in the Appendix. Until the ideal gas law is verified, the Gibbs free energy of $H_2$ is given by:

$$g_{H_2} = g°_{H_2} + k_B T \ln \sqrt{\frac{P}{P°}} \quad (6)$$

The range of validity of this model is however limited by the temperature: phonons are computed in the quasi-harmonic approximation, which limits the range of temperature to below 1000K, after which anharmonic effects should be considered [23].

### 3. Thermodynamic and kinetic models

In the following we present the different steps used to build the global model. In a first step, perfect tungsten is considered in equilibrium with a $H_2$ atmosphere near standard pressure; in such conditions the ideal gas law can be applied and no SAV are formed; a thermodynamic model is built to determine the solubility of hydrogen in these experimental conditions. The model is validated against the experimental data on hydrogen solubility from Frauenfelder [1]. In a second step, single-vacancies are added to the model while an equilibrium with a $H_2$ gas phase is still considered. The total solubility of hydrogen at interstitial sites or trapped in vacancies is determined as a function of the temperature and the pressure. In a third step, a relation between the energy and flux of the impinging ions and the concentration of hydrogen located at interstitial sites is established via a kinetic model. With the concentration of hydrogen at interstitial sites established, we are then able to determine the chemical potential of hydrogen within the range of its implantation depth and consequently, we determine the solubility of hydrogen along with the number of created vacancies.

*3.1 - Hydrogen solubility in perfect W*

We herein consider a tungsten sample in equilibrium with a $H_2$ atmosphere at pressure $P$ and temperature $T$. In a grand-canonical-ensemble picture, the reservoir is the $H_2$ atmosphere containing $N_H$ hydrogen atoms (i.e. $N_H/2$ hydrogen molecules). The system is the tungsten sample in which $n_{int}$ hydrogen are retained at interstitial sites. $N_H$ is so huge as compared to $n_{int}$ that it can be considered as constant. As a consequence, the pressure of the reservoir is constant and the chemical potential it imposes to the system is constant too. On the contrary, $n_{int}$ varies up to the point where equilibrium with the reservoir is established. A schematic representation of such a model is given in the table below.

|             | $H_2$ *(Reservoir)*       | $H_{int}$ *(System – W)* |
|-------------|---------------------------|---------------------------|
| *Initial*     | $\frac{1}{2}N_H$          | 0                         |
| *Equilibrium* | $\frac{1}{2}(N_H - n_{int})$ | $n_{int}$              |

The relevant physical quantity to study such an equilibrium is the Gibbs free energy of the system, which is simply the sum of the Gibbs free energy *per* particle of each constituent of the system *plus* a configurational term:

$$G = \frac{1}{2}(N_H - n_{int})g_{H_2} + n_{int}\, g_{int} - T\, S_{conf} \tag{7}$$

$$S_{conf} = k_B\, ln\left[\frac{\gamma\, N!}{(\gamma N - n_{int})!\, n_{int}!}\right] \tag{8}$$

The configurational entropy expresses all of the configurations that can be built when placing $n_{int}$ hydrogen atom into the *bcc* lattice of a tungsten sample of $N$ W atoms; $\gamma = 6$ is the number of tetrahedral interstitial site in the *bcc* structure of tungsten. The chemical potential µ of H in the reservoir is $\mu = \frac{1}{2}g_{H_2}(P,T)$. The equilibrium conditions being given by $(\frac{\partial G}{\partial n_{int}})_{T,P} = 0$, the atomic fraction $x_{int} = \frac{n_{int}}{N}$ or solubility of hydrogen in tungsten at equilibrium with a reservoir of chemical potential µ is then :

$$x_{int} = \frac{\gamma}{1 + exp\left(\frac{g_{int} - \mu}{k_B T}\right)} \qquad (9)$$

Unless $x_{int} < 10^{-2}$ at.fr., $exp\left(\frac{g_{int} - \mu}{k_B T}\right)$ is large in comparison to 1 and the previous equation simplifies to:

$$x_{int} = \gamma \exp\left(-\frac{g_{int} - \mu}{k_B T}\right) \qquad (10)$$

$$x_{int} = \gamma \sqrt{\frac{P}{P°}} \exp\left(-\frac{g_{int} - \frac{1}{2} g_{H_2}^°}{k_B T}\right) \qquad (11)$$

Equation (11) uses the chemical potential of molecular hydrogen assuming an ideal behavior. It is also known as the Sievert's law, which has been measured experimentally by Frauenfelder [1] for tungsten at high temperature between 1100K and 2400K. In this range of temperature, impurities and other defects are supposed to have only insignificant effects on the solubility itself.

Considering a reservoir at the standard pressure P°, the results from equation (11) are plotted on Figure 1 between room temperature and around 1000K. An extrapolation of the Frauenfelder's law $x_{int} = 9.3 \times 10^{-3} \exp\left(-\frac{1.04}{k_B T}\right)$ at low temperature is provided in Figure 1. The error bars given by Frauenfelder are also used to define the upper and lower boundaries for the solubility within the experimental uncertainty. Using these data, we were able to plot the lower boundary $x_{int}^{B-}$ (or bottom of the error bars) over the full range of temperature according to $x_{int}^{B-} = 5.10 \times 10^{-3} \exp\left(-\frac{1.21}{k_B T}\right)$, and the upper boundary $x_{int}^{B+}$ (or top of the error bars) over the full range of temperature according to $x_{int}^{B+} = 1.82 \times 10^{-2} \exp\left(-\frac{0.87}{k_B T}\right)$.

The agreement between the calculated and experimental solubility is excellent in the range of temperature considered as can be seen from Figure 1. The same model will consequently be used in the next section to determine the hydrogen solubility in tungsten with vacancies.

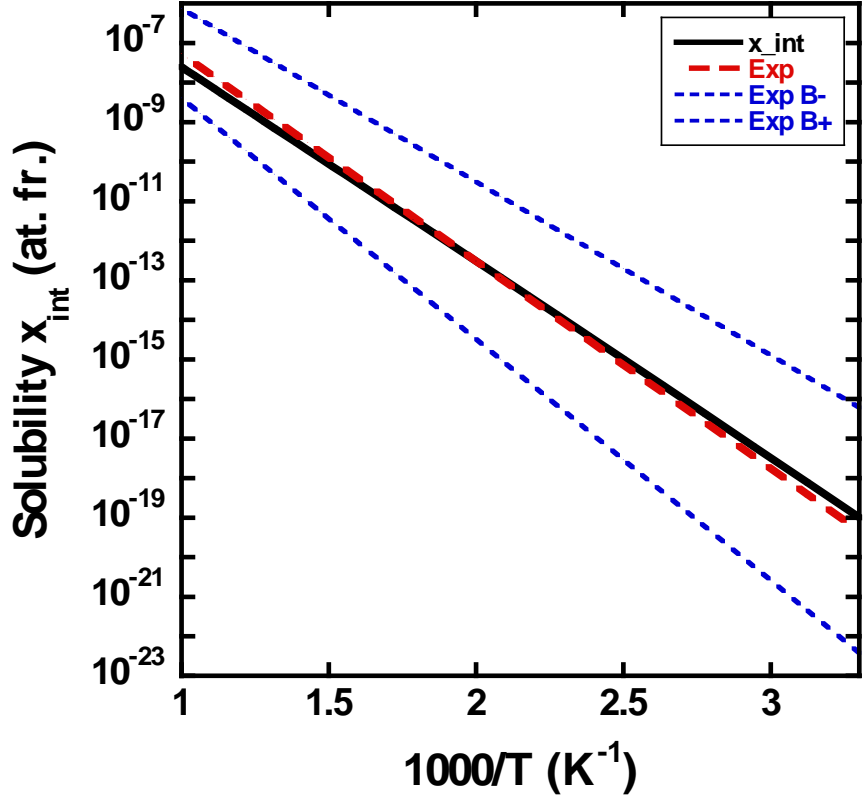

*Figure 1: Calculated solubility $x_{int}$ (black line) compared to the experimental solubility measured by Frauenfleder (red dashed line). Also, the error bars given bar Frauenfelder are plotted (blue dotted lines).*

*3.2 - Hydrogen solubility in tungsten with single vacancies*

A reservoir of $N_H$ atoms in the form of H$_2$ molecules and temperature T and pressure P is again used to impose the chemical potential $\mu = \frac{1}{2} g_{H_2}(P, T)$ to the system. We now consider the case in which the pressure is in the range of GPa. At such a pressure, single-vacancies are created in large number and have to be taken into account into the model. The number of vacancies that trap $j$ H atoms is $n_j$, and the total amount of hydrogen trapped in the single vacancies is $n_{trapped} = \sum_{j=0}^{12} j\, n_j$. The total number of H atoms absorbed in the sample is $n_H$ and the total number of vacancies is $n_v$:

$$n_H = n_{int} + \sum_{j=0}^{12} j\, n_j \qquad (12)$$

$$n_v = \sum_{j=0}^{12} n_j$$

All $n_j$, $n_{int}$, $n_H$ and $n_v$ vary as in the previous paragraph up to the point where the equilibrium is reached. The atomic fraction $x_j$, $x_H$, $x_{int}$ and $x_v$ (same quantities but divided by the number $N$ of W atoms) will be equally used throughout this paper. A schematic representation of such a system is given by the table below.

|  | (Reservoir) | (System – W) | |
| --- | --- | --- | --- |
|  | $H_2$ | $H_{int}$ | $H_{vac}$ |
| Initial | $\frac{1}{2} N_H$ | 0 | 0 |
| Equilibrium | $\frac{1}{2}(N_H - n_H)$ | $n_{int} = n_H - \sum_{j=0}^{12} j\, n_j$ | $\sum_{j=0}^{12} j\, n_j$ |

The Gibbs free energy of the whole system is:

$$G = \frac{1}{2}(N_H - n_H)\mu_{H_2} + n_{int}\, g_{int} + \left(\sum_{j=0}^{12} j\, n_j\, g_j\right) - T\, S_{conf} \qquad (13)$$

$$S_{conf} = k_B \ln Z_{int}^{conf} + k_B \ln Z_{vac}^{conf}$$

$$Z_{int}^{conf} = \frac{\gamma N!}{(\gamma N - n_{int})!\, n_{int}!} \qquad Z_{vac}^{conf} = \frac{(N - n_v)!}{n_v!} \prod_{j=0}^{12} \frac{\omega_j^{n_j}}{n_j!} \qquad (14)$$

$\omega_j$ is the number of degenerate configurations in which $j$ H atoms are located in a single vacancy and was taken from Ref [14]. The equilibrium conditions are now given by $(\frac{\partial G}{\partial n_j})_{T,P,n_i \neq n_j} = 0$ and $(\frac{\partial G}{\partial n_{int}})_{T,P} = 0$.

The first condition leads to:

$$\frac{x_j}{\omega_j(1+x_v)} \left(\frac{\gamma - x_{int}}{x_{int}}\right)^j = \exp\left(-\frac{g_j - j\, g_{int}}{k_B T}\right) \quad (15)$$

while the second condition leads to equation 9 again. Combining equations (9) and (15) gives:

$$\frac{x_j}{1+x_v} = \omega_j \exp\left(-\frac{g_j - j\mu}{k_B T}\right) \quad (16)$$

Noticing that $\sum_{j=0}^{12} x_j = x_v$, one finally gets:

$$\begin{aligned} x_j &= \frac{1}{C}\, \omega_j \exp\left(-\frac{g_j - j\mu}{k_B T}\right) \\ C &= 1 - \sum_{j=0}^{12} \omega_j \exp\left(-\frac{g_j - j\mu}{k_B T}\right) \end{aligned} \quad (17)$$

In cases such as $x_v \ll 1$, then C≈1 and equation (17) takes a very simple analytical form. Equations (9) or (10) and (17) can now be used to determine the atomic fraction of interstitial H atoms and the atomic fraction of H atoms trapped in vacancies, respectively, provided that the chemical potential μ is known. The limit of validity of this model is reached at a concentration around $10^{-2}$ at. fr. (1%) of vacancies. At such a concentration, traps interact with each other, thus the trapping energies for hydrogen are modified as compared to the corresponding trapping energies for diluted vacancies in tungsten. Moreover, aggregation of vacancies and the formation of vacancy clusters is expected. Also, the configurational entropy will take a more complex form as discussed in [13]. This limit is however set at a very high concentration and the domain of validity of the model is consequently quite robust.

*3.3 - Kinetic model*

The objective of this sub-section is to determine the chemical potential μ as a function of the implantation conditions, and consequently all the properties and behavior of hydrogen within tungsten. This can be achieved using equations (10) or (11), provided that the fraction of interstitial H atoms *x*$_{int}$ is known. Based on a simple kinetic model, an analytical expression for *x*$_{int}$ was recently proposed in Ref [3,17].

In this model, an incident flux $\phi_{inc}$ (m$^{-2}$s$^{-1}$) of ions with energy $E_{inc}$ (eV) is assumed to create a triangular depth-profile of interstitial hydrogen atoms in tungsten as represented in Figure 2. The implanted flux is $(1 - r) \cdot \phi_{inc}$ where r is the dimensionless reflection coefficient of the ions and depends on the incident energy $E_{inc}$ of the ions. $R_p$ is the mean depth of ion implantation and is also dependent on $E_{inc}$. The fraction of hydrogen remains maximal at the peak depth over time since the flux of ions continuously implants particles to this depth. This maximal fraction of interstitial hydrogen is called $x_{int}^{MAX}$. The migration of the hydrogen atoms from the implantation zone to the bulk is characterized by the distance $R_d(t)$.

Regarding the profile from $R_p$ to a position which is deeper into the bulk in Figure 2, the linear shape can be justified as follows. If diffusion dominates over trapping, the fraction of diffusive interstitial hydrogen $x_{int}$ at depth $d$ is known [24]; it is $x_{int}(d,t) = x_{int}^{MAX} erfc\left(\frac{d}{2\sqrt{D(T)t}}\right)$, in which $D(T)$ (m$^2$s$^{-1}$) is the diffusion coefficient of interstitial hydrogen. Such a profile is plotted with dashed line in Figure 2. Due to the presence of traps, the *erfc* shape of $x_{int}(x,t)$ is modified and becomes almost linear as shown by macroscopic rate equation and finite element simulations of H in W [25]. This justifies the linear shape we used in our simple kinetic model.

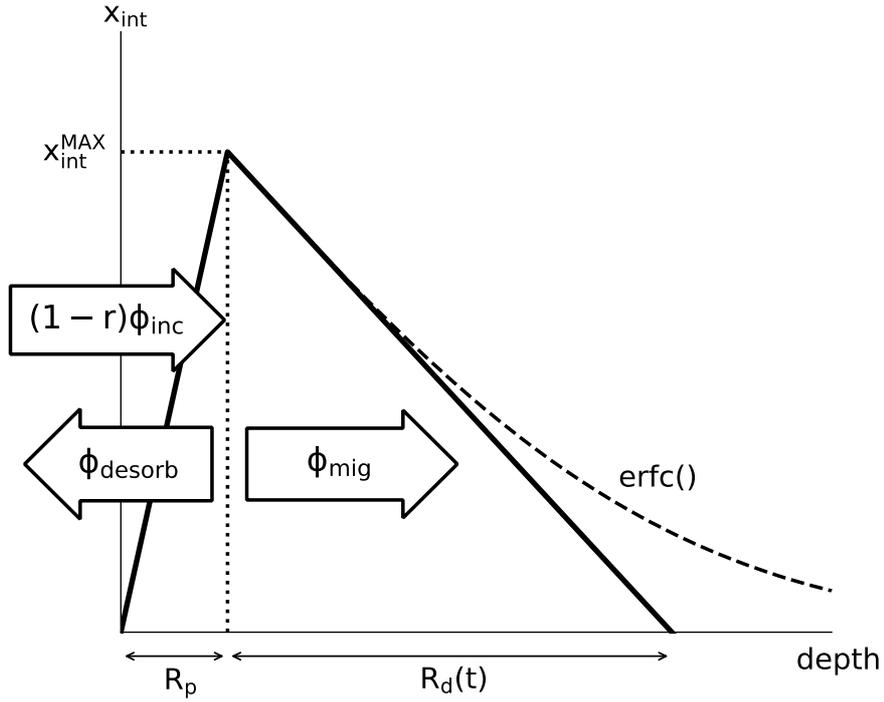

**Figure 2:** *Schematic representation of the depth profile of interstitial hydrogen atoms $x_{int}$ during an implantation with $H^+/D^+$ ions. The profile is shown from the surface (left) to the bulk (right).*

According to Fick's law of diffusion, these two fluxes depend on the diffusion coefficient of hydrogen in tungsten D(T):

$$\phi_{desorb} = D(T) \cdot \frac{x_{int}^{MAX} \cdot \rho_W}{R_p} \quad (18)$$

$$\phi_{mig} = D(T) \cdot \frac{x_{int}^{MAX} \cdot \rho_W}{R_d(t)}$$

Where $\rho_W \approx 6.18 \times 10^{28}$ at. $m^{-3}$ is the tungsten density. The flux balance between the desorbing flux, the migration flux and the implanted flux yields:

$$(1 - r) \cdot \phi_{inc} = -\phi_{desorb} + \phi_{mig} \quad (19)$$

The distance $R_d(t)$ shown in Figure 2 increases with time. $\phi_{mig}$ is inversely proportional to $R_d(t)$; In a semi-infinite material, it consequently decreases and tends towards zero with time. In a real material of finite length $L_0$, $R_d(t)$ will obviously stop once hydrogen reaches the opposite surface, such that $L_0 = R_d + R_p$. The thickness of a W sample is typically around 100 µm. Considering implantation energies of tens to hundreds eV, a typical value for $R_p$ as $R_p \approx 10$ nm, and following equations (18), we have $\frac{\phi_{mig}}{\phi_{desorb}} \approx 10^{-4}$. It means that even though $\phi_{mig}$ does no reach zero in a finite sample, $\phi_{mig}$ is negligible as compare to $\phi_{desorb}$ and can be neglected in equation (19). As a consequence, when the steady state is reached, the implantation and desorption fluxes equilibrate, which leads to:

$$x_{int}^{MAX} = R_p \cdot \frac{(1-r) \cdot \phi_{inc}}{D(T) \cdot \rho_W} \qquad (20)$$

All the quantities are known in equation (20) since $\phi_{inc}$ is given by the experimental conditions, $D(T) = D_o \exp\left(-\frac{E_a}{kT}\right)$ is the diffusion coefficient calculated by DFT [13], and $R_p$ and r can be easily determiner via the binary collision approximation Monte Carlo code SRIM (Stopping and Range of Ions in Matter) [26,27]. SRIM needs $E_{inc}$ as input data, which is also given by the experimental conditions. In Appendix II, the conditions within which a steady state is reached are investigated. Combining equations (9) or (10), (18) and (20) yields the chemical potential of hydrogen, the total solubility of hydrogen and the defect concentration created in the material at Rp, which we examine in the next sub-section.

## 4. Super-saturation in tungsten

*4.1 – Global model*

The kinetic and thermodynamic model are herein combined to yield an estimate of the hydrogen concentration implanted in the sub-surface of tungsten around Rp at a given implantation temperature. The sub-surface is indeed the range of validity of the model; since we use a thermodynamic model, an equilibrium has to be established at least locally. The sub-surface layer is located where the energy of the implanted ions is deposited and thus acts as an energy bath for the system in this region. Consequently, kinetic processes are easily activated in this region, which ensures that equilibrium is reached, and the model applies.

In a first step we model the fraction of hydrogen implanted in the sub-surface. We apply the kinetic model with implantation conditions commonly found in the literature: $\phi_{inc} = 10^{19} m^{-2} s^{-1}$ and $E_{inc}$ = 500 eV/ion. The reflection coefficient r and mean depth of implantation were given by SRIM are r = 0.51 and Rp = 7.7 nm. The kinetic model allows us to determine $x_{int}^{MAX}$ given in equation 20, and then to determine the chemical potential µ from equation 10. The fraction $x_j$ of $VH_j$ vacancies in the sub-surface layers are subsequently calculated using equations (18). These fractions are plotted in Figure 3 in a range of temperature from 300K to 800K.

The most populated vacancies are the $VH_6$ for an implantation at 300K, followed by $VH_7$ and $VH_5$, then $VH_4$ up to $V_0$. This ranking is inverted at higher temperature, namely at 800K. The same qualitative trend exists when an equilibrium with a $H_2$ gas phase is considered [13]; however in the case of implantation, the concentration of vacancies and the solubility of hydrogen dramatically increases to $x_V$ = 0.4% and $x_H$ = 2.4% at 300K. This result is in good agreement with experimental observations that report the formation of super-saturated layers (SSL) within the first ten nanometers of tungsten after implantation at room temperature [3]. It also follows that these SSL would be a consequence of SAV which drastically increase the hydrogen concentration while simultaneously inducing a huge number of defects.

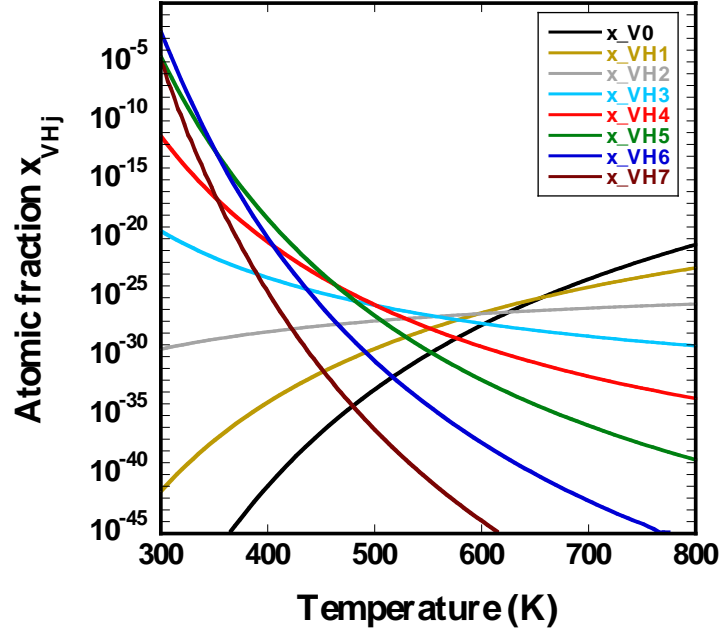

**Figure 3:** *Atomic fraction of VHj vacancies in the sub-surface layer of tungsten within 10 nm with a flux $\phi_{inc} = 10^{19} m^{-2} s^{-1}$ and ion energy $E_{inc}$ = 500 eV/ion.*

To better understand the formation mechanism of the supersaturated layers, we plotted in Figure 4 the total fraction of hydrogen implanted in the sub-surface as a function of the temperature. This was done for fluxes ranging from $\phi_{inc} = 10^{17} m^{-2} s^{-1}$ to $\phi_{inc} = 10^{24} m^{-2} s^{-1}$ and for an incident energy $E_{inc}$ = 500 eV/ion. Two regimes are observed; at high temperatures, the total solubility of hydrogen varies slowly with temperature and remains within one order of magnitude; the concentration is mostly dependent on the flux of implantation. At low temperatures, the total solubility dramatically increases as the temperature decreases; SSL are consequently formed in this regime. The dotted lines in Figure 4 display the fraction of hydrogen trapped at interstitial sites only (i.e. the fraction of hydrogen located within vacancies is removed). It is clear that the dramatic increase in the hydrogen solubility is the consequence of the formation of vacancies in huge numbers that accommodate hydrogen atoms. As a consequence, the physical distinction between these two regimes is driven by trapping at interstitial sites at high temperature, and by the formation of SAV and trapping of hydrogen in the form of VHj vacancies at low temperature. Our model predicts that the formation of SAV and consequent trapping of VHj vacancies in the low temperature regime is what causes the formation of the experimentally observed SSL.

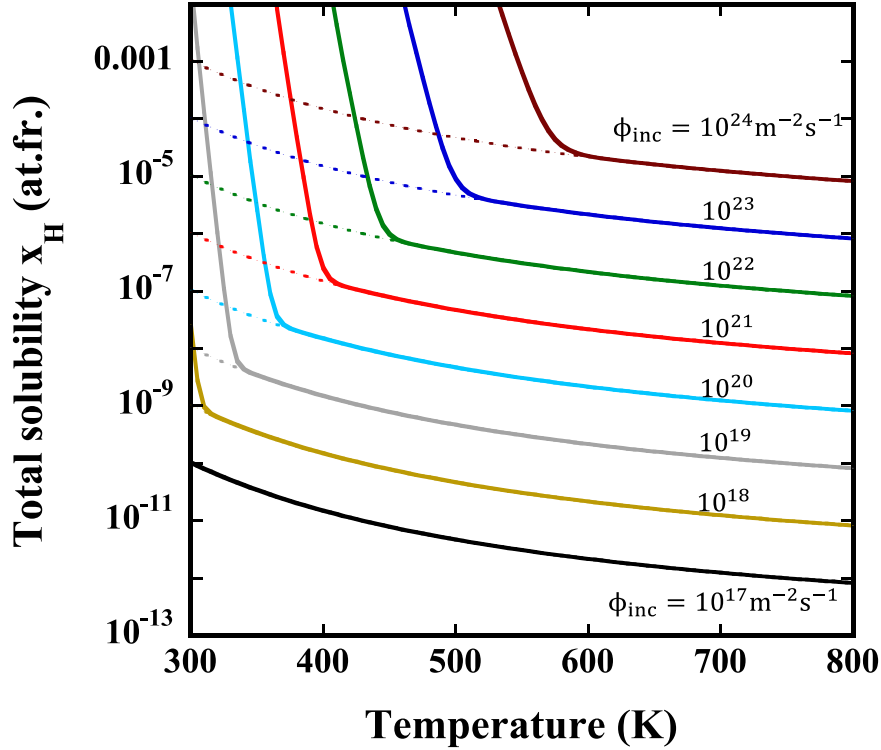

**Figure 4:** *Total solubility of hydrogen (bold lines) implanted at Rp for various incident fluxes ranging from $\phi_{inc} = 10^{17} m^{-2} s^{-1}$ to $\phi_{inc} = 10^{24} m^{-2} s^{-1}$ plotted as a function of the temperature of implantation and for an incident energy of $E_{inc}$ = 500 eV/ion. The fraction of hydrogen trapped at interstitial sites is also plotted in dotted lines for comparison.*

In summary, the temperature of transition between the *vacancy* and *interstitial* regimes depends on the flux: while the *vacancy* regime is not reached at room temperature for a flux of $\phi_{inc} = 10^{17} m^{-2} s^{-1}$, it is reached below 550K with a flux of $\phi_{inc} = 10^{24} m^{-2} s^{-1}$. This temperature of transition $T_t$ is clearly seen in Figure 4 and corresponds to the points where the total solubility and the solubility at interstitial sites are no longer superimposed. Once again, the domain of validity of the present model is in the sub-surface layer where the energy of the implanted ions is deposited. Deeper into the bulk, high concentrations of hydrogen are also measured and were found around $10^{-4}$ *at.fr.* [2,4]. The thermodynamic model remains the same deeper into the bulk, but the formation and diffusion of vacancies are kinetically hindered. The activation energies for diffusion are 1.7eV and 2.1eV for $VH_0$ and $VH_1$, respectively [13].

*4.2 – Temperature of transition depending on the flux*

Below, a simple analytical expression is established for the transition temperature between the *interstitial* and the *vacancy* (or SSL) regimes. At $T_t$, the fraction of hydrogen trapped in interstitial sites $x_{int}$ is equal to the fraction of hydrogen trapped in vacancies. It was already shown [13, 14, 19] that the most populated vacancies at room temperature are $VH_6$; This can be seen in Figure 3 where the population of $VH_6$ dominates by two to three orders of magnitude over $VH_7$ and $VH_5$. This leads to the condition $x_{int} = j\, x_j$ with j = 6.

Equation 10 and 17 with C = 1 allows to determine the transition temperature depending on the chemical potential:

$$T_t = \frac{1}{k_B \ln \frac{\gamma}{j\, \omega_j}} \left[ g_{int} - g_j + (j-1)\mu \right] \qquad (21)$$

The chemical potential is determined using equation (10) and (20):

$$\mu = E_a + g_{int} + k_B\, T_t \ln \left[ \frac{R_p(1-r) \cdot \phi_{inc}}{\gamma\, \rho_W\, D_o} \right] \qquad (22)$$

The transition temperature depending on the flux is the determined combining equations (21) and (22):

$$T_t = \frac{g_j - j\, g_{int} - (j-1)E_a}{k_B \left[ (j-1) \ln \left[ \frac{R_p(1-r) \cdot \phi_{inc}}{\gamma\, \rho_W\, D_o} \right] + \ln \frac{j\, \omega_j}{\gamma} \right]} \qquad (23)$$

$T_t$ can be plotted easily assuming that $g_{int} \approx e_{int} = 1.18$ eV when corrected from the Zero Point Energy (ZPE), $g_j \approx e_j = 4.08$ eV also corrected from the ZPE with j = 6 (using j=5 and $e_5$=3.60ev does not significantly affect the result). The activation energy for diffusion is $E_a$ = 0.20 eV, the other quantities in equation 23 are $D_o = 1.9\ 10^{-7}\ m^2 s^{-1}$, $\gamma = 6$, $\omega_6 = 1$ and $\rho_W$= 6.18 ×$10^{28}$ at.$m^{-3}$. In the end, $Rp$ and $r$ are determined by the incident energy $E_{inc}$ of the ion. They are given by SRIM : r=0.51 and Rp = 7.7 nm at $E_{inc}$ = 500 eV, and r=0.56 and Rp = 2.4 nm at $E_{inc}$ = 70 eV.

$T_t$ is plotted in Figure 6 for $E_{inc}$ equal to 500eV and 70eV. $T_t$ agrees well with the one that can be read on Figure 5 with $E_{inc}$ = 500 eV. As a consequence, equation 23 gives a simple and easy to use analytical expression that allows one to predict the experimental conditions leading to the formation of hydrogen super-saturated layers in tungsten: the SSL are formed in conditions corresponding to the top left of Figure 5, while on the right-hand side no SSL is formed. Figure 6 can be read as a diagram of existence of the SSL.

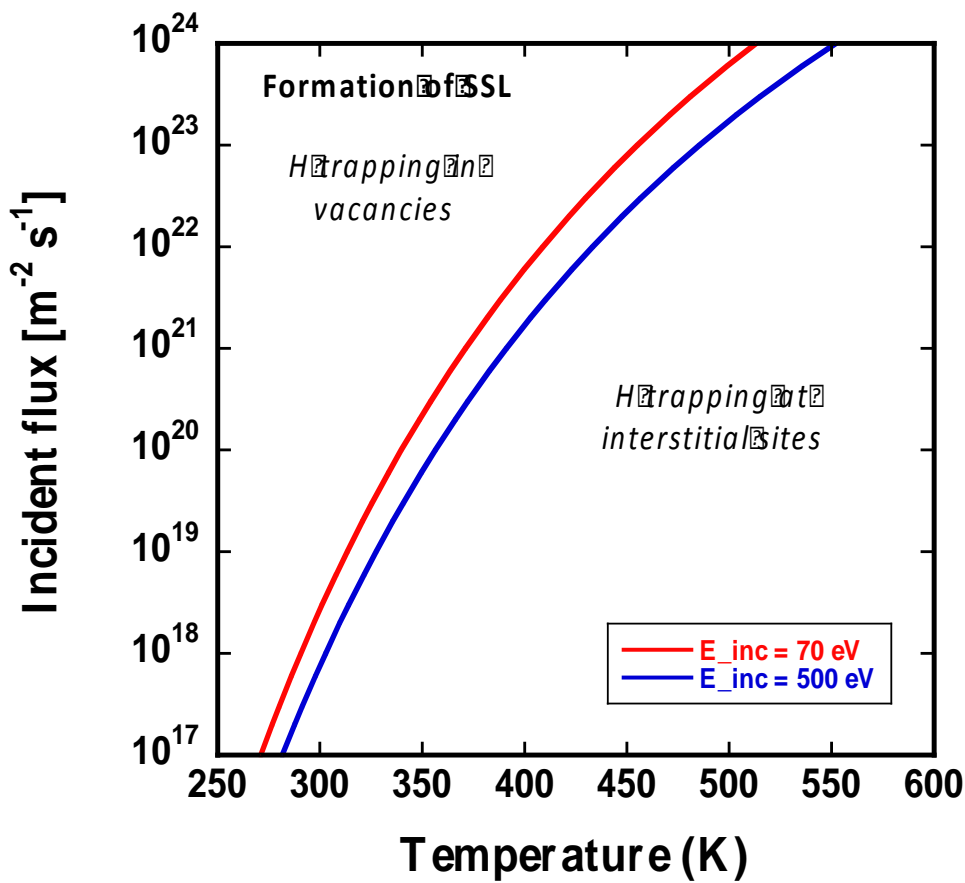

**Figure 5:** *Diagram of existence of the SSL: temperature of transition between the interstitial and vacancy regime for hydrogen trapping plotted as a function of the incident flux of the ions for 2 incident energies, $E_{inc}$=500eV as in Figure 5, and $E_{inc}$ = 70eV.*

Even if this phase diagram still needs to be validated against experimental results, we can nevertheless compare with the results from Gao et al [3] recorded at T=300K with $E_{inc}$ = 70eV and $\phi_{inc}$ = 9.910$^{-19}$ m$^{-2}$s$^{-1}$. With these experimental conditions, they fall in the SSL domain in good correspondence with the concentration experimentally determined to 9.4% *at.fr*. In addition, Poon et al. [28], studied the retention of D in a single crystalline sample implanted by 500 eV/D ions at 300 K and various fluxes; they reported a significant decrease of the deuterium retention for fluxes below 1×10$^{18}$ m$^{-2}$s$^{-1}$; this would indicate that the coordinate T=300K, $\phi_{inc}$ = 1.0 ×10$^{-18}$ m$^{-2}$s$^{-1}$ lies on the transition curve, which is clearly the case in Figure 4 and 5, and would consequently be in full agreement with the model.

## 5. Conclusion

In this work we combined DFT data, a kinetic model, and a thermodynamic model to produce a global model enabling us to examine the behavior of hydrogen in tungsten under hydrogen ions irradiations. The results of our model provide the date to construct a simple diagram allowing one to predict the domain of existence of the super-saturated layers depending on the flux, the energy of the ions, and the temperature of implantation of hydrogen in tungsten. A simple analytical expression is given for the temperature of transition between the *interstitial* and *vacancy* regimes, which make it easy to determine the experimental conditions leading to the formation of the SSL.

**Acknowledgment**

First of all, we would like to express gratitude to our Slovenian and German colleagues for the fruitfull discussions and scientific exchanges we had about the contents of this work:

S. Markelj from JSI, K. Shmid, T. Schwarz-Selinger, A. Manhard, L. Gao and S. Kasper from IPP Garching.

This work was carried out within the framework of the EUROfusion Consortium and received funding from the Euratom research and training programme 2014-2018 under grant agreement No 633053 and also the A*MIDEX project (n° ANR-11-IDEX-0001-02) funded by the 'Investissements d'Avenir' French Government program, managed by the French National Research agency (ANR). The views and opinions expressed herein do not necessarily reflect those of the European Commission. The authors of this work were granted access to the HPC resources of IDRIS and CINES under the allocation A0020806612 made by GENCI (Grand Equipement National de Calcul Intensif) and to the Marconi Supercomputer at CINECA Super Computing Application and Innovation Department, Bologna, Italy.

# Appendix

## Appendix I

The vibrational enthalpies and entropies were calculated as follows:

$$h_{vib} = \sum_{j=1}^{n_{vib}} h\nu_j \left(\frac{1}{2} + \frac{1}{\exp\left(\frac{h\nu_j}{k_B T}\right)-1}\right) \quad \text{(AI-1)}$$

$$s_{vib} = k_B \sum_{j=1}^{n_{vib}} \left[\frac{h\nu_j}{k_B T} \frac{1}{\exp\left(\frac{h\nu_j}{k_B T}\right)-1} - \ln\left(1 - \exp\left(-\frac{h\nu_j}{k_B T}\right)\right)\right] \quad \text{(AI-2)}$$

For the gas phase, it is also necessary to considered the translational et rotational components to the Gibbs free energy. Since H$_2$ is an homonuclear diatomic molecule, these components were computed as follows:

$$h_{trans} = \frac{5}{2} k_B T \qquad s^\circ_{trans} = k_B \left(\frac{5}{2} + \ln\left[\frac{k_B T}{P^\circ}\left(\frac{2\pi m k_B T}{h^2}\right)^{\frac{3}{2}}\right]\right) \quad \text{(AI-3)}$$

$$h_{rot} = k_B T \qquad s_{rot} = k_B \left(1 + \ln\left[\frac{8\pi^2 I k_B T}{\sigma h^2}\right]\right) \quad \text{(AI-4)}$$

In $s^{rot}$, σ depends on the symmetry of the molecule. It is 2 for an homonuclear diatomic molecule. *I* is the inertia momentum of H$_2$.

## Appendix AII

We herein check the conditions within which the steady state leading to equation [15] is reached. To this end, we assume the concentration of trapped hydrogen is constant up to the depth $R_d(t)$. This concentration of trapped hydrogen is called $x_{trapped}$. It follows the number of hydrogen trapped per surface unit is $c_{trapped} = \rho_W R_d(t) x_{trapped}$. The flux which is feeding the growth of this quantity is $\phi_{mig}$. It came that, $\phi_{mig} = \frac{d\,c_{trapped}}{dt}$. Taking the expression of $\phi_{mig}$ from equation (18), this leads to $R_d(t)\,dR_d(t) = \frac{D(T)}{x_{trapped}} x_{int}^{MAX}\,dt$, which is $R_d(t) = \sqrt{\frac{2D(T)x_{int}^{MAX}}{x_{trapped}} t}$ once integrated over time. Using this expression in the flux balance (13)

$$0 = \frac{D(T)}{R_p} \cdot \left(\sqrt{x_{H_i}^{MAX}}\right)^2 + \sqrt{\frac{x_{H_{trapped}} \cdot D(T)}{2 \cdot t}} \cdot \sqrt{x_{H_i}^{MAX}} - (1-r) \cdot \frac{\phi_{inc}}{\rho_W} \quad \text{(AII-1)}$$

we end up with:

$$\sqrt{x_{int}^{MAX}} = \sqrt{R_p \cdot \frac{(1-r)\cdot\phi_{in}}{D(T)\cdot\rho_W}} \cdot \sqrt{\frac{\tau_m}{t}} \cdot \left(\sqrt{1+\frac{t}{\tau_m}} - 1\right) \quad \text{(AII-3)}$$

Where $\tau_m = \frac{R_p \rho_W x_{trapped}}{8(1-r)\phi_{inc}}$ is the time characterizing the growth of $x_{int}^{MAX}$. Figure A shows the evolution of $x_{int}^{MAX}/x_{int}^{MAX}{}_{t\to\infty}$ as a function of $t/\tau_m$ as described by equation (14). It requires $t = 360\tau_m$ for $x_{int}^{MAX}$ to reach 90% of its steady-state value defined by the equation (20) which can be considered as the steady-state condition. In experiment, a typical implantation flux is $(1-r)\phi_{inc} = 10^{19}\,Dm^{-2}s^{-1}$ with ion energy of 250 eV/D. This ion energy, at normal incidence, leads to $R_p = 5\,nm$. Considering $10^{-2}$ *at.fr* as an upper limit for $x_{trapped}$, one can obtain $\tau_m = 0.0425\,s$. Thus, the steady-state condition is $t > 15.3\,s$. Considering the experimental runs last for hundreds or even thousands of seconds to reach fluence > 1021 Dm-2, it appears very reasonable to assume that the steady-state is reached during experiments.

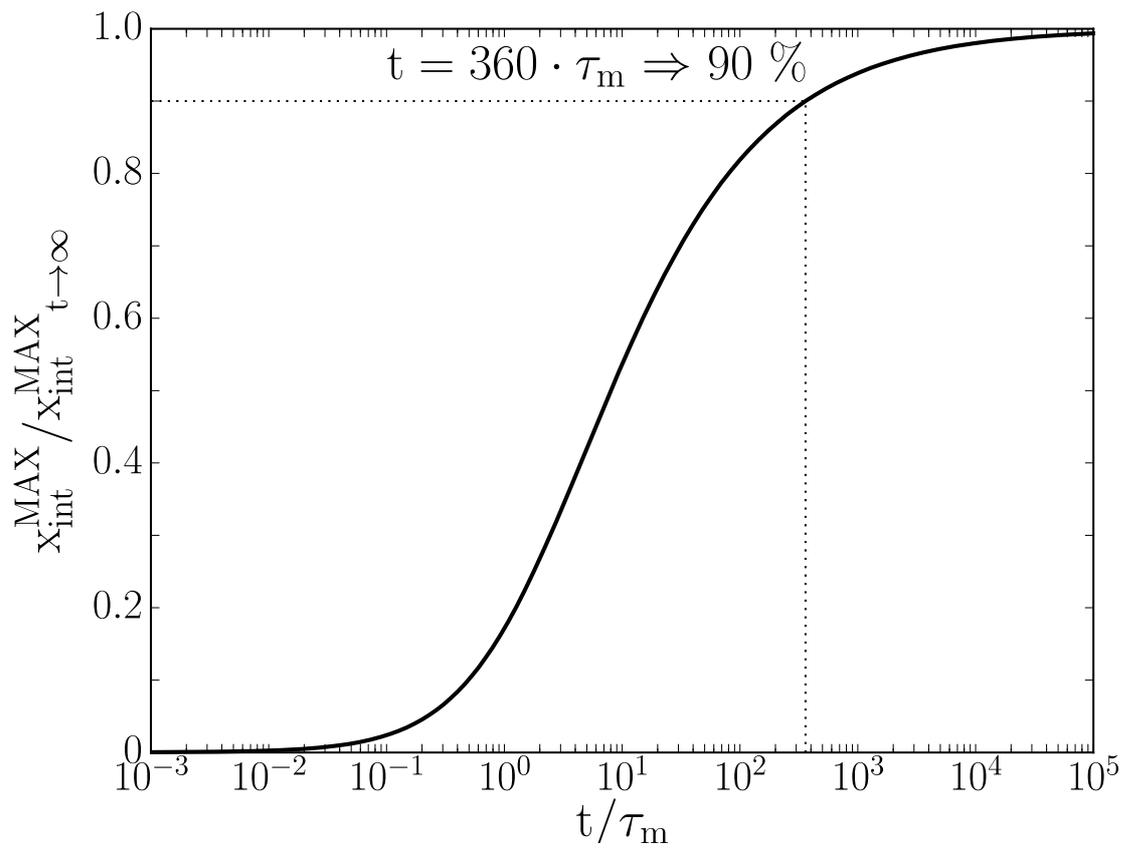

**Figure A:** evolution of $x_{int}^{MAX}$ with time $t/\tau_m$.

temperature dependence of D retention in single crystal tungsten J. of Nucl. Mater. 313-316 (2003) 199